\documentclass[aps,prb,showpacs,twocolumn,amssymb,floats,epsfig]{revtex4}
\usepackage{graphicx}
\usepackage{amsmath}
\usepackage{mathbbol}
\usepackage{relsize}
\usepackage{afterpage}
\usepackage{braket}
\usepackage{hyperref}
\usepackage{subfigure}
\usepackage{adjustbox}
\usepackage{float}
\usepackage{graphicx}

\newcommand{\ct}{\cite}

\newcommand{\be}{\begin{equation}}
\newcommand{\ee}{\end{equation}}
\newcommand{\ba}{\begin{eqnarray}}
\newcommand{\ea}{\end{eqnarray}}

\usepackage[title,titletoc,toc]{appendix}

\usepackage[usenames,dvipsnames]{color}

\begin{document}

\title{Quenching in Chern insulators with satellite Dirac points: The fate of edge states}
\author{Utso Bhattacharya, Joanna Hutchinson and Amit Dutta}
\affiliation{\small{Department of Physics, Indian Institute of Technology, Kanpur 208016, India}}

\begin{abstract}
We perform a sudden quench on the Haldane model with long range interactions, more specifically generalising  to the next to next nearest neighbour hopping, referred to as the $N3$ model in our
work. Such a model possesses both isotropic and multiple anisotropic (satellite) Dirac points which lead to a rich topological phase diagram consisting of phases with  higher Chern number ($C$). Quenches between the topological and the non-topological phases of such an infinite system probe the effect of the presence of the anisotropic Dirac points on the non-equilibrium response of the topological system. Interestingly,  the Chern number remains the same before and after the quench for both the quenching protocols, even when the quench of the system is carried out between two different topological phases.  {However, for a finite system, we establish that the initial edge current asymptotically decays to zero when the system is
quenched to the non-topological phase although the Chern number for the corresponding periodically wrapped system remains unaltered; what is remarkable is that when the Hamiltonian is quenched from $|C|=2$ phase to the non-topological phase the edge current
associated with the inner channel  decays at a faster rate than the outer channel resembling a situation in which the system passes through the phase with $|C|=1$ before ending up in the phase $C=0$.  }
\end{abstract}


\maketitle

 \section{Introduction}
 \label{intro}

{The ability to classify and understand  different phases of matter happens to be the cornerstone of condensed matter physics. Many situations, for
example,   superconducting and magnetic transitions (both classical and quantum),  can be explained solely on the basis of the celebrated Landau theory of spontaneous symmetry breaking
\cite{chaikin95,cardy,sachdev,duttaetal}.  However, the advent of new phenomena like Integer Quantum Hall effect (IQHE)  and Fractional Quantum Hall effect (FQHE) have revealed the inadequacy of such a theory in explaining the quantization of Hall conductance in systems where no spontaneous symmetry breaking occurs \cite{goerbig09}. This subtle new order in the pattern of the ground state entanglement has necessitated the reformulation of the electronic band theory in the form of topological band theory, so that, not only the dispersion relation of the bands but also the non-trivial evolution of the eigenvectors encapsulating the inherent topological order of the system is more apparent \cite{h,hk,Qi10,moore,Qi11,bernvig13}.}

{The Quantum Hall effect (QHE) mentioned above is a phenomenon which happens when the current flowing through a 2D electron gas maintained at a very low temperature is subjected to a very strong perpendicular magnetic field; this results in the transverse (Hall) conductance getting quantized in units of some integers or some special fractions times a universal constant \cite{goerbig09}. Although, IQHE can easily be understood within the framework of single particle dynamics using topological band theory,
it took quite some time to realize that it is the breaking of the time reversal symmetry (TRS) that plays a key role in getting the transverse (Hall) conductance quantized. Haldane in his seminal paper \cite{h} utilized this realization by putting forward a different mechanism that can explicitly break TRS exhibiting quantum anomalous hall effect (QAHE). This is achieved with the use of a staggered magnetic field, producing chiral edge states whilst maintaining zero total flux through each plaquette. This model on a 2D honeycomb lattice (with real nearest and complex next nearest neighbour electron hopping), popularly known as the Haldane model, is an ideal example of a Chern insulator featuring topologically distinct phases of matter and brings to fore a mechanism through which a quantum Hall effect can appear as an intrinsic property of a band structure, effected by the breaking of both TRS and inversion symmetry (IS), rather than being effectuated by an external magnetic field. The Haldane model and its topological phase diagram has of course, now been experimentally realised using ultra cold fermionic atoms in a periodically modulated optical honeycomb lattice \cite{jotzu14}. }

{The concept proposed by Haldane eventually
 gave birth to a new class of materials called the topological insulators which are characterized by non-trivial topological order \cite{hk,Qi11}.  This material  behaves as  insulator in its bulk, but possess conducting electronic states on its surface, thereby, exhibiting quantum-Hall-like behaviour in the absence of a magnetic field. Topological systems are thus, part of a richer class of systems, thereby, playing a crucial role in our understanding of quantum phase transitions \cite{duttaetal}, quantum fidelity \cite{garnerone09, mukherjee12},  Loschmidt echo and  quantum decoherence \cite{Patel13_marg}, and also quantum quenches \cite{bermudez09,Patel13,rajak14,sacra1,ccb,privitera15, sacramento16}. In particular, Chern insulators demonstrating nontrivial (Chern) phases, serve as an invaluable theoretical tool to investigate these novel phases of matter, and further our understanding of topological band theories. Furthermore, these materials can be viewed as building blocks for other symmetry protected insulators  with  robust  edge  states,  for  example  quantum spin  hall  insulators \cite{kane05,bernevig06,zhangetal}, topological superconductors which host Majorana fermions\cite{fu08}  and 3D  topological  insulators \cite{fkm} which are also promising candidates for fault tolerant quantum computation due to the presence of symmetry protected surface states robust against back-scattering \cite{kitaev09}.}

{
Given the recent development of experimental
studies especially in cold atoms and optical lattices and possible rich technological applications,
understanding the non-equilibrium dynamics  of closed quantum systems under the application of quantum quenches \cite{calabrese,PolkovnikovRev,dziarmaga10}, periodic drives \cite{bdp,inouetan,oka09,kitagawa10} 
and  the dynamics of open systems coupled to external reservoirs\cite{dom1,dom2}, is the most challenging problem at present. The quenched systems are also connected
to dynamical phase transitions \cite{heyl13,sharma15,sharma16}, work-statistics \cite{gambassi11,russomanno15} and emergent thermodynamics \cite{dorner12,sharma15_PRE}. Furthermore,
periodically driven system have gained importance from the viewpoint of Floquet graphene \cite{oka09,kitagawa10}, Floquet topological insulators \cite{lrg} (for review see \cite{cdsm}) and
dynamical generation of Majorana bound states \cite{thakurathi13,thakurathi14};
these studies are also important from the aspects of defect generation \cite{mukherjee09,das10}, dynamics of decoherence \cite{quan06,sharma12,mukherjee12a,nag12}, dynamical saturation \cite{ russomanno12,sharma14}, dynamical localization \cite{nag14,agarwala16} as well as many body localization \cite{lazarides15}.

To investigate the non-equilibrium behavior  (e.g., the subsequent unitary evolution and the relaxation behavior)  following a quench, one of the parameters of the Hamiltonian describing the system  is tuned suddenly or slowly (with respect to the inverse of the energy gap in the system). 
 Recently, much attention has been devoted to study the effect of a sudden quench of the parameters of a Hamiltonian with topological properties and, particularly, in such a way so that the new Hamiltonian ends up in a different topological phase or at the critical point separating them \cite{bermudez09,Patel13}. It is then of vital importance to observe how the robustness of the topological properties and edge states in finite sized systems respond with time to such changes. It has recently been shown that the toric code models \cite{hh} and the topological superconductors \cite{sacra1} are quite resilient to such sudden quenches. 
quenches studied in the present work.
Recently, Caio et al\cite{ccb} has investigated the non-equilibrium response of the Chern insulator by performing a sudden as well as  a slow quench on the Haldane model.  It was shown that the Chern number of the initial ground state remains unchanged throughout the post quench unitary evolution  for an infinite Haldane system modelled on a hexagonal lattice, for both the quenching protocols involved, no matter in which topological phase the final Hamiltonian ends in;  { this point was also illustrated in Ref. [\onlinecite{alessio15}]. We would like to emphasize that 
 the  preservation of the winding number of the many-body state following a quench was
already noted in Ref. [\onlinecite{fdgy1, fdgy2}]; these works
focus on quenches in interacting topological BCS superfluids, but
the asymptotic long-time behavior can be described by an effective
single-particle Bogoliubov-de Gennes  Hamiltonian.}

These remarkable results motivate us to further investigate in this paper the fact whether this observation happens to be an artefact of the special nature of their selected model. The Haldane model based on a Graphene like lattice with electron hopping only up to nearest neighbour hosts two inequivalent Dirac points within the first Brillouin zone, with a linear, isotropic dispersion relation (i.e., they possess the same velocity along both the perpendicular directions). We therefore address the question: what will happen if the same kind of quenching (both slow and sudden) is performed on a Hamiltonian which has Dirac cones with deformed conical dispersions centred on each of them? A deformed conical dispersion can originate when the Dirac points are anisotropic which means that they have different velocities along the two perpendicular directions. In short,  the aim of this work is to understand how this modified  nature of dispersion affects the Chern index of the initial ground state of the unquenched Hamiltonian in any topological phase after a sudden or slow quench to the same or some other topological phase.}

\begin{figure}
\centering
\subfigure[]{%
\includegraphics[width=0.6\textwidth,height=7cm]{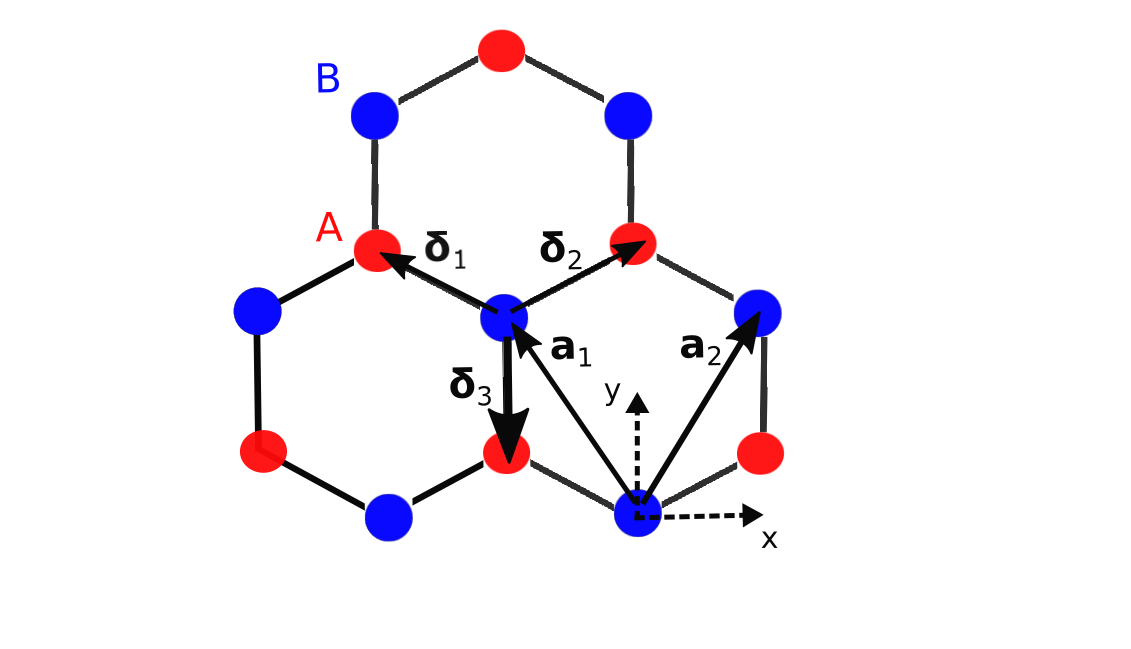}
\label{fig:hex1}}
\quad
\subfigure[]{%
\includegraphics[width=0.35\textwidth,height=5cm]{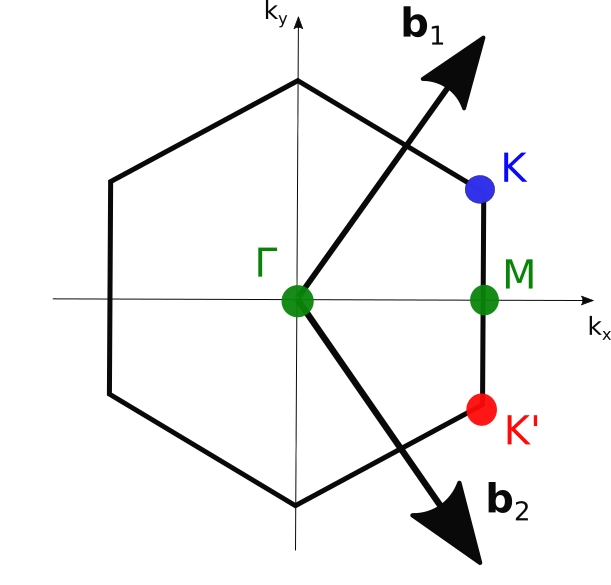}
\label{fig:hex2}}
\caption{(Color online) (a) Three real space plaquettes of the hexagonal lattice with lattice vectors $\mathbf{a}_1$ and $\mathbf{a}_2$. The  blue and red circles represent the two sublattices A and B.
(b) BZ for hexagonal lattice with reciprocal  lattice vectors $\mathbf{b}_1$ and $\mathbf{b}_2$ with $K$ and $K'$ representing two inequivalent Dirac points. }
\label{fig.Dirac_merge_contour}
\end{figure}

{Until recently, theoretical studies of Chern insulators, e.g., the Haldane model of graphene, has been limited to models that are based on nearest ($N1$) and next nearest neighbour interactions ($N2$) only, producing topological phases which are labelled by Chern numbers taking the values of $\pm 1, 0$. In a recent work, Sticlet \textit {et al.}, \cite{sp}, established that it is indeed possible to obtain large Chern number phases by introducing longer range of electron hopping, as 
first discussed in \cite{bs}, leading to an increase in the number of Dirac points (DP's) and hence, the conductance. A first non-trivial generalisation of the nearest neighbour Haldane model along these lines would be to introduce a next to next nearest neighbour hopping ($N3$). This new model was found to host three further Dirac points encircling each of the two isotropic Dirac points seen earlier (when the electron hopping was only restricted to $N2$). But, very promisingly, these new DP's also known as the satellite Dirac points (SDP's) have remarkable features: Unlike the original DPs of the $N2$ model, these SDPs of the $N3$ model, have an anisotropic but linear dispersion relation (a deformed conical dispersion centred on the SDP's). This  deformed conical dispersion
enables us to understand the role  of the anisotropy when the $N3$  model is subjected to the above mentioned quenches. Of course, this model as we will enunciate  later has a topological phase diagram with higher number of Chern phases labelled by the Chern numbers $\pm 2, \pm 1, 0$. Hence, it would also be immensely interesting to see how the presence of these new topological phases affect the initial ground state after the quenches.}

 {According to the bulk-boundary correspondence, the fact that the periodically wrapped system is in the higher Chern phase should be reflected in the existence of equivalent number of  edge channels (in the system with an open
boundary condition) in the equilibrium situation; however,  to the best of our knowledge, the nature of edge states for a  $N3$ Haldane model has not been presented in an explicit form in earlier studies \ct{bs,sp}. We therefore rigorously establish  the presence of edge states in the equilibrium situation. Further, we address the question what happens to the edge current following  a quench; this is never a priori obvious from the notion of the conservation of Chern numbers. 
 Remarkably, our study  establishes that  if  the pre-quench equilibrium Hamiltonian has a Chern index $|C|=2$,  the current associated with the outer  edge states decays at a slower rate than the inner
edge states before the edge current finally vanishes;  this behaviour of the edge current, as if, mimics a situation in which the system initially  is in $|C|=2$ phase (like two edge channels in the open system in equilibrium)  reaches the $C=0$ (no edge state) phase through an intermediate
$|C|=1$ (single edge state) phase. Thus, even though the Chern index remain invariant, the consequence of the quench on the open system in  a higher Haldane phase is reflected in the decay of
pre-quench edge current; an intermediate $C=1$ like behavior is manifested in the difference between the decay rates of two channels with the current in the inner channel decaying faster
than the outer channel.}

 {The paper is organised in the following way:  we begin with a summary of the {equilibrium} $N3$ model in Sec. \ref{model}, where we also provide a thorough discussion on the nature of the edge states choosing a finite geometry.   We then perform a sudden quench in Sec. \ref{quench_gen}  and discuss the post-quench value of the  Chern number.   On the other hand,  the behaviour  of the initial edge current following a quench from the topological to the non-topological phase is summarised  in Sec. \ref{sec_edge_dyn}.   Finally we present the concluding comments   in Sec.  \ref{conc}. We have
also included two appendices which supplement the results presented in the main text.}

\section{The model}
\label{model}

\begin{figure}
\begin{center}
\includegraphics[width=0.4\textwidth,height=0.2\textheight]{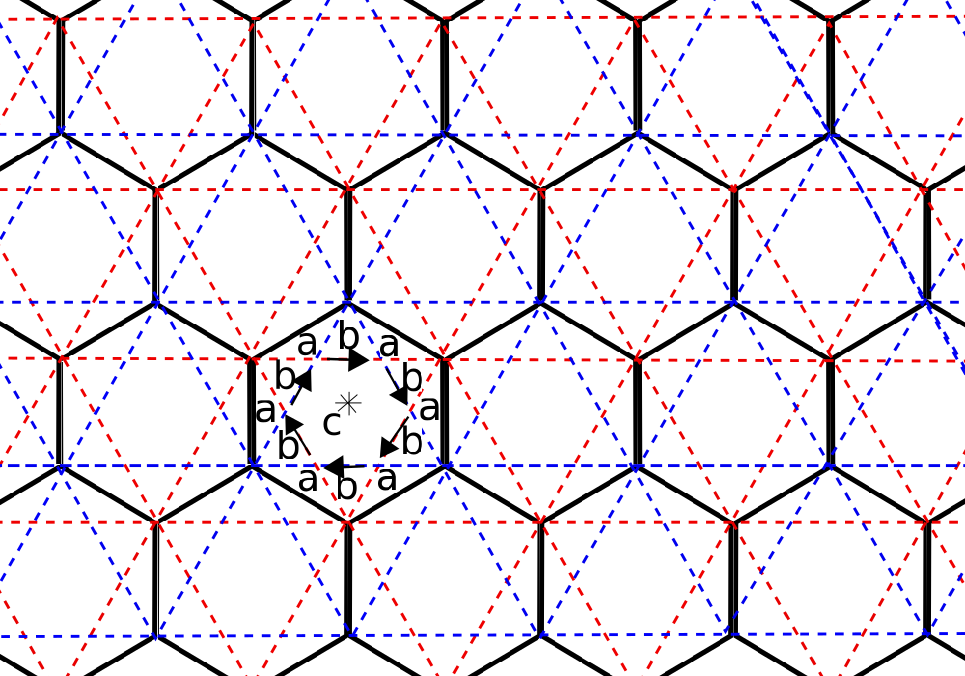}
\end{center}
\caption{(Color online) (Color online) A hexagonal lattice, where nearest neighbour interactions are shown in bold and next nearest neighbour by the (dashed) blue and red triangles  for the two sublattices A and B. The regions of the BZ labelled $a(b)$ are regions through which the flux is $\phi_{a(b)}$. }
\label{latt}
\end{figure}

We consider a 2D model on a hexagonal lattice comprised of two triangular sublattices $A$ and $B$ as shown in Fig. \ref{fig.Dirac_merge_contour}. It can be understood as the composition of $N3$ graphene  {in the presence of the sublattice symmetry (SLS) breaking mass term ($M$)  and  a staggered magnetic field which manifests itself
in the complex next nearest hopping}, with Hamiltonian 
\begin{equation}
\begin{aligned}
&\mathcal{H} =  t_1 \sum_{i,j=N1}  \left(c^\dagger_{iA} c_{jB} + h.c. \right) \\
&+  t_2 \sum_{i,j=N2}  \left( e^{i\phi_{ij}} \left(c^\dagger_{iA} c_{jA} + c^\dagger_{iB} c_{jB} \right) + h.c. \right) \\
&+ t_3 \sum_{i,j=N3}  \left( c^\dagger_{iA} c_{jB}   + h.c. \right) \\
&+ M \sum_{i\in A} \hat{n}_i - M \sum_{i\in B} \hat{n}_i.
\label{eq_haldane_ext}
\end{aligned}
\end{equation}
The $c_{iA(B)}$s are spinless fermionic operators on sublattice $A$ ($B$), and the $t_i$s are the $i^{th}$ nearest neighbour hopping interaction strengths. The TRS of this model is broken by the phase factor $\phi_{ij}=\pm \phi$, originating from the staggered magnetic field and is positive for anticlockwise nearest neighbour hopping (see Fig.~\ref{latt}).

Fourier transforming into k-space the Hamiltonian becomes
\begin{equation}
\begin{aligned}
\mathcal{H} & = \begin{pmatrix}
c^{\dagger}_A(\mathbf{k}) & c^{\dagger}_B(\mathbf{k}) 
\end{pmatrix}
\mathbf{h} (\mathbf{k})
\begin{pmatrix}
c_A(\mathbf{k}) \\
c_B(\mathbf{k}) 
\end{pmatrix},
\end{aligned}
\end{equation}
where 
\begin{equation}
\mathbf{h}(\mathbf{k}) = \sum^3_{i=0} h_i (\mathbf{k}) \sigma_i.
\label{eq_ham}
\end{equation}
The $\sigma_{i}$, $i \in \left\{ 1,2,3\right\}$ are the Pauli matrices, $\sigma_0$ is the identity matrix and $a$ is the lattice constant. 
We have 
\begin{align}
h_0 &= 2 t_2 \cos (\phi) \biggl[\cos \left( \mathbf{k} \cdot \mathbf{a}_1 \right) + \cos \left( \mathbf{k} \cdot \mathbf{a}_2 \right) \nonumber \\
&\hspace{4mm} + \cos \left( \mathbf{k} \cdot \left(\mathbf{a}_1 - \mathbf{a}_2 \right)\right) \biggr], \\
h_1 &= t_1 \biggl[ 1 + \cos \left( \mathbf{k} \cdot \mathbf{a}_1\right) + \cos \left(\mathbf{k} \cdot  \mathbf{a}_2 \right) \biggr]\nonumber\\
&\hspace{4mm}+ t_3  \biggl[ \cos \left(\mathbf{k} \cdot \left(\mathbf{a}_1 + \mathbf{a}_2\right) \right)+ 2 \cos \left(\mathbf{k}\cdot \left(\mathbf{a}_1 - \mathbf{a}_2\right) \right) \biggr],\\
h_2  &= t_1   \biggl[\sin  \left( \mathbf{k} \cdot \mathbf{a}_1 \right) +\sin \left( \mathbf{k} \cdot\mathbf{a}_2 \right) \biggr] - 2 t_3 \sin \left( \mathbf{k} \cdot \left(\mathbf{a}_1+\mathbf{a}_2 \right) \right),\\
h_3 & = M + M_H, \\
M_H &= 2 t_2 \sin (\phi) \biggl[\sin  \left(\mathbf{k} \cdot \mathbf{a}_2 \right)  - \sin \left( \mathbf{k} \cdot \mathbf{a}_1 \right) \nonumber\\
&\hspace{4mm}+ \sin \left( \mathbf{k} \cdot \left(\mathbf{a}_1 - \mathbf{a}_2 \right) \right)\biggr],
\end{align}
where {$M_H$ is the Haldane mass that vanishes for $t_2=0$ or $\phi=0$} and $\mathbf{a}_1 =\frac{a}{2} \left(\sqrt{3},3 \right)$ and $\mathbf{a}_2 =\frac{a}{2} \left(- \sqrt{3},3 \right) $ as shown in Figure \ref{fig:hex1}.

\begin{figure}
\begin{center}
\includegraphics[width=0.4\textwidth,height=0.2\textheight]{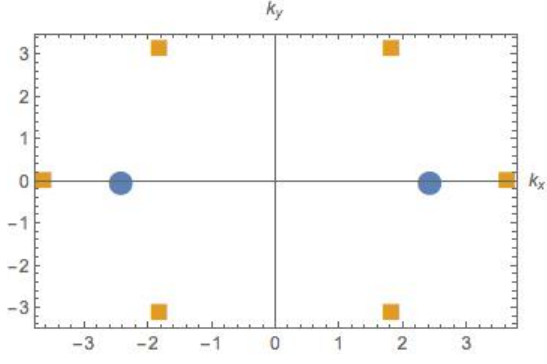}
\end{center}
\caption{(Color online) (Color online) A schematic representation of the orange (square) anisotropic SDPs encircling the blue (circular) isotropic DPs in k-space, with $t_1=3t_3 =a=1$.
The anisotropic DPs associated with an isotropic DP are related to each other by C$_3$ symmetry.}
\label{dps}
\end{figure}

When the system is restricted to nearest neighbour hopping, for $M= \phi=0$, the two bands touch at the six corners of the hexagonal Brillouin zone (BZ). Only two of these are inequivalent and are time-reversed partners of each other.
However, as found in \cite{bs}, increasing the interaction range, gives the possibility for obtaining further DPs. In particular, when the interaction range is increased to third-nearest-neighbour ($N3$), the positions of the original isotropic DPs, protected by TRS, SLS and C$_3$ symmetry remain unchanged. [C$_3$ symmetry is an inherent symmetry in the bare Graphene lattice born out of the fact that the electron hopping matrix elements must be invariant upon the cyclic permutation of the nearest neighbor hopping (bond) vectors (as shown in Figure \ref{fig:hex1}) $\delta_1\implies\delta_2\implies\delta_3\implies\delta_1$ is called  the C$_3$ symmetry. This symmetry can easily be understood by noticing that the lattice structure remains unchanged after a rotation of $2\pi/3$ around the center of the hexagonal plaquette or around any of sites of the two sublattices denoted by A and B. We must however note that although this symmetry constrains the nearest neighbour hopping terms to be the same, the $n$-th next nearest hopping terms need not be equal. This implies that only cyclic permutations (and not all permutations) of the nearest neighbour hopping vectors obey the C$_3$ symmetry.]   Additionally we find three SDPs encircling each of the the original DPs, for $t_3$ in the range
\begin{equation}
t_3 \in [ -\infty , - t_1 ]\cup [ \frac{t_1}{3}, \infty].
\end{equation}
These SDPs are given by the $\mathbf{k}$ value for which $h_1(\mathbf{k}) = h_2( \mathbf{k})=0$ and determined by  the C$_3$ symmetry  of the lattice  as explicitly
shown in Eq. \eqref{eq_aniso_dirac} (see also Fig.~\ref{dps}). Unlike the original (isotropic) DPs, these additional SDPs have different velocities along the two orthogonal directions in k-space as
we shall show explicitly in Eq.\eqref{defns}, and hence are anisotropic.

{
For the sake of the completeness we quote below  the co-ordinates of both isotropic and anisotropic DPs  in k-space as illustrated in Fig.~\ref{dps}:
 For the two inequivalent isotropic DPs, these are given by
\begin{equation}
\mathbf{k}^{\alpha} = \alpha K \left( 1, 0 \right), \qquad K = \frac{4 \pi}{3 a \sqrt{3}}, 
\end{equation}
for $\alpha= \pm1$, where $a$ is the lattice constant.
The other $4$ isotropic DPs can be obtained by C$_3$ rotation of these points.
 {The coordinates for the $6$ anisotropic SDPs are given by}
\begin{equation}
\begin{aligned}
\mathbf{k}^{\alpha,n} & = \alpha k \left( \cos \frac{2 (n-1) \pi}{3}, \sin \frac{2 (n-1) \pi}{3} \right), \\
k & = \frac{2}{a\sqrt{3}} \arccos \left(\frac{t_3 - t_1}{2 t_3} \right),
\label{eq_aniso_dirac}
\end{aligned}
\end{equation} 
where $n=1,2,3$ corresponds to the $3$ anisotropic DPs encircling each isotropic DP ($\alpha=\pm1$).
Throughout the rest of this paper we will use the variable $\alpha=\pm$ to denote the two isotropic DPs, and $n \in \left\{1,2,3\right\}$ to refer to the satellite (anisotropic) DPs associated with each of the isotropic points. }

\subsection{In the continuum limit}

The low-energy description at half filling for $N3$ graphene with the TRS and SLS breaking mass terms, is given by the sum of 8 Hamiltonians expanded about each of these 8 DPs (2 isotropic DPs and 6 anisotropic SDPs). In each case Eq.~\eqref{eq_ham} has the form
\begin{equation}
\mathbf{h}^{\alpha,n} (\mathbf{k}) = \begin{pmatrix}
M + M_H^{\alpha,n} (\phi) &  |\mathbf{k}^{\alpha,n}|  \bar{z}^{\alpha,n} \\
 |\mathbf{k}^{\alpha,n}| z^{\alpha,n} & - \left( M + M_H^{\alpha,n} (\phi)  \right) \\ 
\end{pmatrix},
\label{hiso}
\end{equation}
with
\begin{equation}
\begin{aligned}
& h^{\alpha,n}_{1} = \alpha c_x \delta k^{\alpha,n}_{x}, \qquad h^{\alpha,n}_2 = c_y   \delta k^{\alpha,n}_{y},\\
& h^{\alpha,n}_3  = M+ M_H^{\alpha,n} (\phi);\\
\end{aligned}
\end{equation}
Here, $c_x$ and $c_y$ are the Fermi velocities of the Dirac particles and 
\begin{equation}
\begin{aligned}
& \delta k^{\alpha,n}_{x} = | \mathbf{k}^{\alpha,n} | \cos \theta, \qquad  \delta k^{\alpha,n}_{y}   = | \mathbf{k}^{\alpha,n} | \sin \theta, \\
\end{aligned}
\end{equation} 
\begin{equation}
z^{\alpha} = \alpha c_x \cos \theta -i c_y \sin \theta,
\label{z}
 \end{equation}
 where $\theta$ measures the angular deviation of any point from the $x$-axis in $k$-space measured from any of the eight gapless points (DPs or SDPs).
 
 For the 2 isotropic DPs we have 
 \begin{equation}
 \begin{aligned}
 &c_x=c_y=- c, \qquad c = 3a \left( \frac{t_1}{2}  -   t_3 \right), \\
 &M_H^{\alpha} =  -3 \sqrt{3} \alpha t_2 \sin \phi.
 \label{isodefs}
 \end{aligned}
\end{equation}
 On the other hand, for the 6 anisotropic SDPs we instead have
{
\begin{equation}
\begin{aligned}
& c_{x} = 2 \sqrt{3}  \sqrt{ 1- \left(\frac{t_{3} - t_{1}}{2t_{3}} \right)^2} c, 
c_{y} = \frac{3a}{2} \left(\frac{2t^{2}_{3}-t^{2}_{1} + t_{1}t_{3}}{t_{3}} \right),\\
& M_H^{\alpha} (\phi) = 
 \quad - 2 \alpha \left( \frac{t_{2}}{t_{3}} \right) \left( t_{3} + t_{1} \right) \sqrt{1-\left( \frac{t_{3}-t_{1}}{2t_{3}} \right)^{2}} \sin \phi. \\
\label{defns}
\end{aligned}
\end{equation}}

It is worth noticing that the mass term $M_H^{\alpha,n}(\phi)$ is the same for each SDP (i.e., independent of $n$), but different from that for the isotropic DPs (i.e. depends
on $\alpha$)..
 We will make this distinction clear throughout the rest of this paper by relabelling
\begin{equation}
M+ M_H^{\alpha} (\phi) = 
\begin{cases}
\mathcal{M}^{\alpha} (M, \phi) & \text{for isotropic DPs,} \\
m^{\alpha} (M, \phi) & \text{for anisotropic SDPs.} \\
\end{cases}
\label{masses}
\end{equation}
 
 
The eigenstates for the Hamiltonian expanded about the anisotropic SDPs are given by 
\begin{equation}
\ket{\psi^{\pm}_{n,\alpha}} = 
\begin{pmatrix}
\mp \frac{\bar{z}}{|z|} f_{\pm} \\
f_{\mp}
\end{pmatrix}, 
\end{equation}
where
{
\begin{equation}
f_{\pm} \left( \mathbf{k}^{n, \alpha}, m^{\alpha},\theta \right)= \sqrt{\frac{1}{2} \left( 1 \pm \frac{m^{\alpha} }{E^{\alpha,n}} \right)},
\label{f}
\end{equation}
with corresponding eigenvalues, $\lambda_{\pm}  = \pm E^{\alpha,n}$,
 \begin{equation}
 \begin{aligned}
 E^{\alpha,n}  \left( \mathbf{k}^{n, \alpha}, m^{\alpha},\theta \right)  = \sqrt{(m^{\alpha} )^{2}+ (\mathbf{k}^{\alpha,n})^2 | z^{\alpha,n} |^{2} }.
 \label{evals}
 \end{aligned}
\end{equation}}

The eigenstates for the isotropic DPs can be similarly obtained by using the values for $c_x$, $c_y$ and $M^{\alpha}_H$ given in Eq. \eqref{isodefs} (thus replacing $m^\alpha$ with $\mathcal{M}^\alpha$). In the above analysis, we have neglected the $h_0$ term since it does not effect the eigenstates and thus the topological properties (or the direct gap) of the system, and we have set $\hbar =1$ throughout. 

\subsection{Pre-quench Chern number}
\label{cnb}

One can now compute the Chern number \cite{sp} when the energy spectrum is nonvanishing, which requires that either $\phi$ or $M$  is nonzero (presuming the system has DPs where the energy bands touch when the mass term $h_3( \mathbf{k})=0$). In particular, a topological phase  will only be achieved if time-reversal symmetry is broken ($\phi \neq 0$) \cite{h}. 

Since, we are interested to see how the Chern number of the ground state of the pre-quench Hamiltonian changes with time after the quench, we cannot just use the projectors of the final Hamiltonian. It then becomes unavoidable to use the following prescription to compute the Chern number for the state of the low-energy system $\ket{\psi}$, by integrating the Berry curvature $\Omega $ over 2D momentum \cite{bernvig13} space to obtain,
\begin{equation}
\nu = \frac{1}{2\pi} \int^{2 \pi}_0  d \theta \int^\infty_0 dk \text{ } \Omega,
\label{cno}
\end{equation}
where  $\Omega = \partial_k A_\theta - \partial_\theta A_k$ and $A_x = i\bra{\psi} \partial_x \ket{\psi};x=k,\theta$. 

\begin{figure}
\begin{center}
\includegraphics[width=0.35\textwidth,height=0.25\textheight]{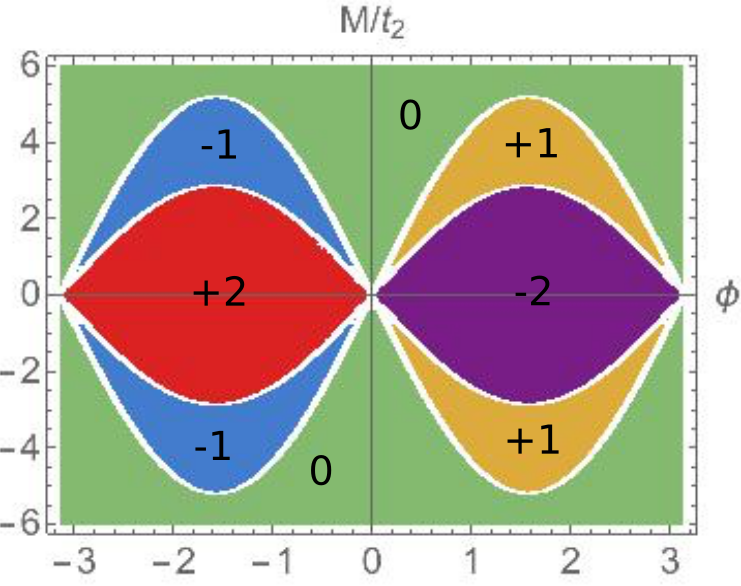}
\end{center}
\caption{(Color online) (Color online) Chern number phase diagram for the Haldane model on N3 graphene, plotted as function of the ratio of the on-site energy $M$ to the next nearest neighbour hopping term $t_2$ along the vertical axis against the staggered flux $\phi$ along the horizontal axis, with $t_1=1$, $t_2=1/3$ and $t_3=0.35$. The outer lines of transition separates the trivial insulating phase ($C=0$) from the non-trivial ($C=\pm 1$) Chern phases, whereas the inner lines of transition are the exclusive features of a higher Chern phase model. The Chern number actually changes by three units from $\pm 1$ to $\mp 2$ across these transition lines due to the creation of the three SDPs.}
\label{chernno}
\end{figure}

Introducing a TRS and SLS breaking mass term ($M$ and $M_H$, respectively) for $N1$ graphene, one can induce topological phases characterized by  Chern numbers taking values $\left\{-1,0,1\right\}$: For $N3$ graphene this is increased to include $\pm 2$ (as detailed in Appendix \ref{prequapp}). 
We note that either the $\mathcal{M}^{\alpha}(M, \phi)$ or $m^{\alpha}(M, \phi)$ term will change sign at the boundaries between the different phases, shown in Fig.~ \ref{chernno}.

The contribution to the Chern number from each SDP for the ground state of this system is found to be 
\begin{equation}
\nu^{\alpha,n} = - \frac{1}{4 \pi} \text{sgn}(m^{\alpha})  \int^{2 \pi}_0 \frac{\partial \delta}{\partial \theta} d \theta,
\label{chern}
\end{equation}
where
\begin{equation}
\delta = - \alpha \arctan \left( \frac{c_y}{c_x} \tan \theta \right),
\label{delta}
\end{equation}
and for the isotropic DPs $\nu^{\alpha} = -\frac{\alpha}{2} \text{sgn}(\mathcal{M}^{\alpha})$. See Appendix \ref{prequapp} for details.

The total Chern number for a state of the system is the sum of the contribution from each DP;
\begin{equation}
\begin{aligned}
\nu = \frac{1}{2} & \left(  \text{sgn } \mathcal{M}^- - \text{sgn } \mathcal{M}^+ \right) \\
& - \frac{3}{2} \left( \text{sgn } m^- - \text{sgn } m^+ \right).
\end{aligned}
\end{equation}

\subsection{Edge currents}

\label{sec_edge}

Let us now investigate the  edge states in the extended Haldane model \eqref{eq_haldane_ext}   {choosing a finite geometry  periodic in  the $x$-direction and with a finite width of length $N$} in the $y$-direction with arm chair edges;  {we bridge a connection with Chern numbers derived in the previous section 
and establish 
that the results are in congruence with the bulk-edge correspondence.
}

\begin{figure}[H]
\centering
\subfigure[]{%
\includegraphics[width=0.35\textwidth,height=3.9cm]{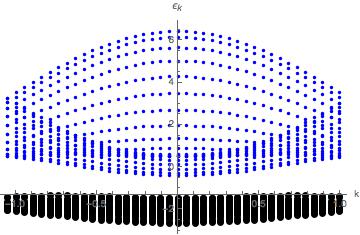}
\label{fig:espec0}}
\quad
\subfigure[]{%
\includegraphics[width=0.35\textwidth,height=3.9cm]{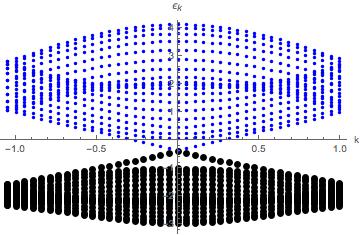}
\label{fig:espec1}}
\quad
\subfigure[]{%
\includegraphics[width=0.35\textwidth,height=3.9cm]{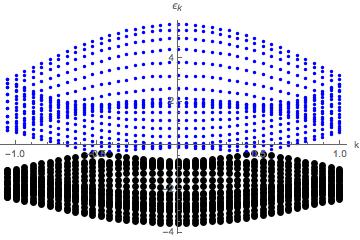}
\label{fig:espec2}}
\caption{(Color online) The energy spectrum of the Hamiltonian in Eq.~ \eqref{eq_haldane_ext} with parameters $t_1=1$, $t_2={1}/{3}$, $M = 1$, and $N=20$;  the  blue  (small)  dots and black 
(big) dots  represent the conduction band and  the valence band, respectively. In Fig.~(a)  the parameters $t_3=0.5$ and $\phi = {\pi}/{12}$, correspond to the non-topological phase in Fig.~
\ref{chernno} with  $C= 0$ and hence there is  no crossing of energy bands.  In Fig.~(b), we choose  $t_3=0$ and $\phi = {\pi}/{3}$, so that the periodically wrapped system is in the
phase $C=1$,  consequently there is one band  crossing at the centre (implying the existence  of edge modes).  Finally,  in Fig.~(c)  $t_3=0.5$ and $\phi = {\pi}/{3}$, so that  $C=-2$; the band crossing at the centre disappears while two
new crossing emerge at the corner.}
\label{espec}
\end{figure}

In Fig. \ref{espec},  {we present the energy spectrum for different phases  {by choosing three set of parameter values which
correspond to three distinct phases  in the phase diagram in Fig. \ref{chernno}, with Chern number
$C=0$, $1$ and $-2$, respectively; this} clearly shows a correspondence between the Chern number and the number of band crossings. In the non-topological phase, $C=0$, there is no band  crossings (Fig. \ref{fig:espec0}) i.e., there is no zero-energy edge states. On the contrary, as shown in Figs. \ref {fig:espec1}  for $C=1$,   a single crossing is
obsereved. Finally, when   the parameters are tuned
such that the system is in the higher Chern phase with $C=2$, the band crossing at the centre disappears while two new band crossings emerge at the corners (Fig.~\ref{fig:espec2}).}

We now define a local current flowing through site $i$ by
\begin{equation}
\hat{\mathbf{J}}_i = - \frac{i}{2} \sum_j \mathbf{\delta}_{ji} \left(t_{ij} \hat{c}^\dagger_i \hat{c}_j - h.c. \right), 
\label{eq_curr_op}
\end{equation}
where $t_{ij}$ represents the hopping parameter between sites $i$ and $j$, $\mathbf{\delta}_{ji}$ is the vector displacement of site $i$ from $j$, and the sum is over nearest, next nearest and next to next nearest neighbour interactions.
Noting that each site can be labelled by $(x,y,s)$, where $x$ and $y$ represent the position and $s$ indicates whether the site is in sublattice $A$ or $B$, we can  derive  the total longitudinal current flowing along the strip in the $x$-direction at a definite transverse $y$-position through the relation,
\begin{equation}
J_x = \langle\hat{J}_x \rangle = \sum_{y,s} \langle \hat{J}_{x,y,s} \rangle.
\label{eq_edge_curr}
\end{equation} 
The $\langle . \rangle$ represents the expectation value  over the current carrying the ground state of the Hamiltonian).

\begin{figure}[H]
\centering
\includegraphics[width=0.35\textwidth,height=4.5cm]{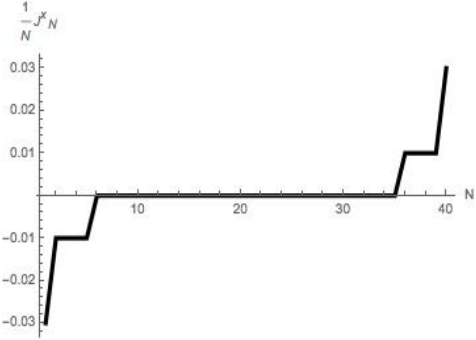}
\caption{(Color online) The average current in the $x$-direction along each row $N=1,\ldots, 40$ for $t_1=1$, $t_2={1}/{3}$, $t_3=0.5$, $\phi = {\pi}/{3}$, and $M=0$. We clearly see
the existence of two edge channels. The bulk current as expected, however, stays at zero.}
\label{jperrow}
\end{figure}

Figure \ref{jperrow} presents the average current in the $x$-direction along each row in the phase with $C=-2$;  the current is nonzero in both the  edge channels on each side and  flows in the same direction along both edge channels on the same side. Remarkably, the outer channel carries  a  current  of higher magnitude than the inner channel. As expected, the bulk current vanishes.

\section{Sudden Quench  and the  post-quench Chern number}

\label{quench_gen}


We begin by preparing the system at time $t=0$ in the ground state $\ket{\psi^-_{n,\alpha}}$ of the Hamiltonian  with parameters $(M,\phi)$. Next we perform a sudden quench, abruptly changing the parameters to new values $(M',\phi')$, and allow the system to evolve unitarily under the action of this new Hamiltonian.

Immediately after quenching, the pre-quench ground state of the system can be written as the following superposition of the eigenstates of the post quench Hamiltonian,
{
\begin{equation}
\begin{aligned}
\ket{\psi^-_{n,\alpha}(k,\theta)}  = & a_{n, \alpha} (k, \theta) \ket{\psi'^l_{n,\alpha}(k,\theta)} \\
& + b_{n, \alpha} (k, \theta) \ket{\psi'^u_{n,\alpha}(k,\theta)},
\label{gs}
\end{aligned}
\end{equation}}
where $\ket{\psi'^{l(u)}}$ are the new lower (upper) post-quench eigenstates.
\label{cna}
{The unitary time evolution of the system is now governed by the post quench Hamiltonian parametrised by $(M',\phi')$. Thus, the time evolved state of the system at an instant of time $t$ after the sudden quench, can be written as,
\begin{equation}
\begin{aligned}
\ket{ \psi_{n,\alpha}(k,\theta, t)} = & e^{-ih^{\alpha,n}(k, \theta, M', \phi')t}\ket{\psi^-_{n,\alpha}(k,\theta)} \\
 = & e^{i E' t} a_{n, \alpha} (k, \theta) \ket{\psi'^l_{n,\alpha}(k,\theta)} \\
& + e^{-i E' t}   b_{n, \alpha} (k, \theta) \ket{\psi'^u_{n,\alpha}(k,\theta)}. \\
\end{aligned}
\label{timeev}
\end{equation}}
{where, $\pm E'$, are the eigenvalues of the post quench Hamiltonian $h^{\alpha,n}(k, \theta, M', \phi')$ having the same form as in Eq.~(\ref{hiso}) but with parameters ($M', \phi'$) in place of ($M, \phi$). The expansion coeffecients $a_{n, \alpha} (k, \theta)$ and $b_{n, \alpha} (k, \theta)$ depend on the $m^\alpha(M,\phi)$, $m^\alpha(M',\phi')$ and the energy eigenvalues (see Eq. (\ref{evals})) of both the pre and post quench Hamiltonians and can easily be obtained by equating both the sides of Eq.~(\ref{gs}):
\begin{equation}
\begin{aligned}
a_{n, \alpha} & = f_- f'_- + f_+ f'_+ , \\
b_{n, \alpha} & = f_+ f'_- - f_- f'_+,
\end{aligned}
\label{eqn_ab}
\end{equation}}
{where Eq.(\eqref{f}) clearly specifies the form of $f_{\pm}=f_{\pm} \bigl( \mathbf{k}^{n, \alpha}, m^{\alpha}(M, \phi),\theta \bigr)$ and $f'_{\pm}=f_{\pm} \bigl( \mathbf{k}^{n, \alpha}, m^{\alpha}(M', \phi'),\theta \bigr)$ in Eq.~(\ref{eqn_ab}) above.}\\
{Computing the Chern number for the satellite DPs of this new post-quench state of the system, we find that the coefficients {$a_{n, \alpha}$ and $b_{n, \alpha}$ along with $A_k$ are $2 \pi$ periodic in $\theta$; the $\theta$ dependence only enters through the $z$ dependence (see Eq.\eqref{z}) of the energy eigenvalues (Eqs.~\eqref{evals}) appearing in the $f, f'$} as defined through Eq. \eqref{f}. The post-quench Chern number  given in Eq. \eqref{nu} of Appendix \ref{prequapp},  in this case gets simplified to
\begin{equation}
\nu(t) = \frac{1}{2\pi} \int^{2 \pi}_{0} \frac{\partial \delta}{\partial \theta} \left[h (k,\theta) \right]^\infty_0 d \theta.  
\label{nu2}
\end{equation}}

{The function $h(k,\theta)$ is given by 
\begin{equation}
\begin{aligned}
h(k, \theta) & = \frac{a_{n, \alpha}b_{n, \alpha} k |z| }{E'}   \cos 2 E' t - \left( a_{n, \alpha}^2 f'^2_- + b_{n, \alpha}^2 f'^2_+\right),
\label{h}
\end{aligned}
\end{equation}
has the following limits :
\begin{equation}
\begin{aligned}
& h(\infty,\theta) = - \frac{1}{2},  \\
& h(0,\theta) = - \frac{1}{2} \left( 1 - \text{sgn}(m^\alpha) \right).
\end{aligned}
\end{equation}
We thus recover Eq.\eqref{chern}, and find the Chern number contribution from the anistropic DPs remain unchanged after the quench.}

{Using the same arguments given in Appendix \ref{prequapp} we can recover the Chern number contribution from the isotropic DPs, which match the results of Caio \textit{et al} \cite{ccb}, thus concluding that the total Chern number remains unaltered post-quench. Again, we refer the reader to Appendix \ref{postquapp} for details of the calculation.}

\begin{figure}
\begin{center}
\includegraphics[width=0.35\textwidth,height=0.2\textheight]{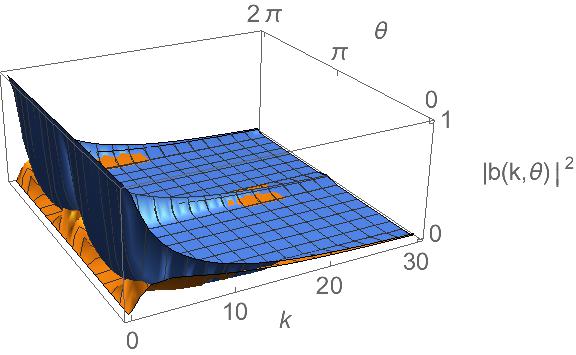}
\end{center}
\caption{(Color online) (Color online) The probability of occupying the excited state $|b_{n, \alpha}(k,\theta)|^2$ for a single anisotropic SDP, with $t_1=1$ and $t_3 =0.35$ after a quench of $m^{\alpha}$ for a sign-preserving quench from $-1$ to $-0.1$ (yellow surface) and for a sign-changing quench from $-1$ to $0.1$ (blue surface).}
\label{bgraph}
\end{figure}

Chern numbers can be understood as a winding number about each of the zeros of the energy spectrum (the DPs). For a Hamiltonian expanded about each of these points, the Chern number is the winding around the position $\mathbf{k}=0$, where $\mathbf{k}$ is the distance from the DP. A sudden quench performed between two topologically different phases causing a sign change in either $\mathcal{M}^{\alpha}$ or $m^{\alpha}$, will give {$|b_{n, \alpha}(0,\theta)|=1$} (see Eq. \eqref{eq_ab}), meaning the post-quench system is in the excited state. Such a system in an excited state can be thought of as one in the ground state with a mass $\mathcal{M}^{\alpha}$ or $m^{\alpha}$ of opposite sign. Thus we conclude that quenching between topologically different phases leaves the total Chern number unchanged. In the situation where $\mathcal{M}^{\alpha}$ and $m^{\alpha}$ have not changed after the quench, we find that { $b_{n, \alpha}(0,\theta)=0$}, thus the system remains in the ground state and it is expected that the Chern number should not change. {Hence, it becomes apparent that the $\theta$ dependence of the DPs can only be visualised as a local deformation of the energy landscape in the BZ which does not affect the global topological properties of the infinite two level system.}

\section{Dynamics of edge states following the quench}

\label{sec_edge_dyn}

 {Finally, we examine the role of dynamics on the behaviour of these edge currents presented in Sec. \ref{sec_edge}. We achieve this by observing the behaviour of the edge currents after performing a sudden quantum quench choosing the parameters in such a way that the system is quenched from the phase with $C=-2$ to the non-topological phase. It is noteworthy that to study the temporal evolution of the edge current, we need to calculate the expectation value of the current operator defined in Eq. ~\eqref{eq_curr_op} over
the quenched state, (i. e., the time evolved counterpart  of  the initial state evolved with the time independent final Hamiltonian)  in the expression \eqref{eq_edge_curr} (not just over the ground state as in the equilibrium case).}

\begin{figure}[H]
\centering
\quad
\subfigure[]{%
\includegraphics[width=0.35\textwidth,height=4.5cm]{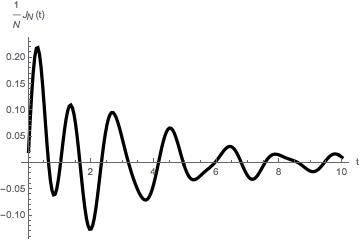}
\label{fig:dyn1}}
\quad
\subfigure[]{%
\includegraphics[width=0.35\textwidth,height=4.5cm]{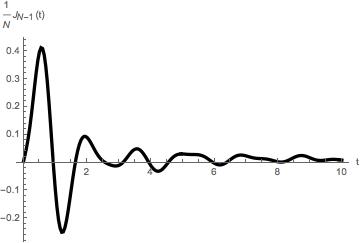}
\label{fig:dyn2}}
\caption{(Color online) The average current in the $x$-direction plotted as a function of time, for $t_1=1$, $t_2={1}{/3}$, $t_3=0.5$, $\phi = {\pi}/{3}$, and $N=40$, following a quench  from the
mass value  $M={1}/{3}$ (corresponding to a topological phase with $C=-2$) to $M=2$  (non-topological phase, $C=0$)  along the rows  $N$ (Fig.~(a)) and  $N-1$ (Fig.~(b).)}
\label{dyn}
\end{figure}

 {Analysing results presented in Fig. \ref{dyn}, we find that (at $t=0$) the current  along both the edges states starts from their pre-quench values (as given in Fig.  \ref{jperrow}). Interestingly, the initial 
current  vanishes in the asymptotic limit (for $t \gg 0$) following  an oscillatory pattern.  Remarkably,  despite the fact that the Chern number remains unchanged after the quench the edge current
asymptotically assumes the value of the non-topological phase where edge states do not exist as shown in Fig. \ref{fig:espec0}.}

 {Furthermore, what needs to  be emphasised is that the current associated with the inner edge channel (Fig. \ref{fig:dyn2}) decays faster than the outer edge channel (Fig.~\ref{fig:dyn1}). Thus at an intermediate time, we see that only the outer edge channel is current carrying; it therefore resembles a system in a topological phase with (bulk) $C=-1$.  If the pre-quench Hamiltonian is in the higher Chern phase with   $C=\pm 2$, after quenching to a non-topological phase with $C=0$,  we therefore encounter a fascinating  situation:  the system  {\it apparently} passes through a topological phase with $C=\pm 1$ before ending in one with Chern number $0$.}

\section{Conclusion}
\label{conc}

 {We study the Haldane Hamiltonian in the presence of the next to next neighbour hopping; in such a long-range interacting model,  there exist satellite Dirac points which are of anisotropic  nature.}
In this paper, we have explored the quenching dynamics of such an extended Haldane model when one of the parameters (i.e., the mass term of the 
Hamiltonian) is quenched suddenly  to ananlyze the role of such additional DPs on the quenching.  Remarkably,   our study reveals that 
the quenching between topologically different phases leaves the total Chern number unchanged. In the situation where $\mathcal{M}^{\alpha}$ and $m^{\alpha}$ have not changed after the quench, we find that { $b_{n, \alpha}(0,\theta)=0$}, thus the system remains in the ground state and it is expected that the Chern number should not change. {We conclude that the $\theta$ dependence, which can be considered as a local deformation of the energy landscape in the BZ, does not affect the  {the preservation   of the Chern number which is thus robust against introducing higher range interactions leading to additional DPs. However, more fascinating
situation emerges when we probe the same system with an open boundary condition in the context of the dynamics of the edge current. }

 {Our study analyses the edge channels of the equilibrium $N3$ Hamiltonian considering a finite geometry and establishes the existence of the two edge channels (one outer and
the other inner) for the phase $|C|=2$ as expected from the bulk-boundary correspondence.  Interestingly, the magnitude of current is higher for the outer channel. 
 Remarkably, following the quench to the non-topological phase, the initial edge current vanishes in the asymptotic
limit; this implies  that the existence of the edge current in the asymptotic limit is the property of the post-quench Hamiltonian which in equilibrium does not support topological edge states. Furthermore,
the current associated with the inner edge channel decay faster resembling  a situation in which the system passes through the phase with $|C|=1$ before ending up in the phase $C=0$. This is
remarkable given that no bulk-boundary correspondence exists in the non-equilibrium situation.}

Given that the Haldane model has been experimentally realized, one
expects that quenching experiments can be performed and predictions made in this paper are viable to experimental verifications.


\textbf{Acknowledgements: }JH gratefully acknowledges support from the Leverhulme Trust during this work. AD acknowledges SERB, DST, New Delhi
for financial support.

\begin{appendices}

%


\section{Chern number before the quench}
\label{prequapp}

We begin by computing the Chern number for the Hamiltonian expanded about a Dirac point for an initial state $\ket{\psi_{n,\alpha}} $ of this two level system, given by a superposition of the lower (upper) eigenstates {$\ket{\psi^{-(+)}}$;
\begin{equation}
\begin{aligned}
\ket{\psi_{n,\alpha}(k,\theta)} = & a(k, \theta) \ket{\psi^-_{n,\alpha}(k,\theta)} \\
& + b(k, \theta) \ket{\psi^+_{n,\alpha}(k,\theta)}.
\end{aligned}
\end{equation}}
In general we assume that the coefficients $a$ and $b$ will depend on both $k$ and $\theta$; the parameters defining the BZ. However when considering the expansion of the system about an isotropic DP, we find them to be independent of $\theta$.

{
We compute the Chern number according to Eq.\eqref{cno} and find for the anisotropic SDPs:
\begin{equation}
\begin{aligned}
 A_k   =  & i \frac{m |z| }{2 E^2 } \left( a^* b  - b^* a  \right) 
, \\
A_\theta = & i \frac{m}{2 E k |z|} \frac{\partial E}{\partial \theta} \left( a^* b - b^* a \right) \\
& +\frac{1}{2}   \frac{\partial \delta}{\partial \theta}   \frac{k |z|}{ E} \left( a^* b + b^* a \right)  \\
& -\frac{1}{2}   \frac{\partial \delta}{\partial \theta} \left( |a|^2  \left( 1 - \frac{m}{E} \right)  + |b|^2   \left( 1 + \frac{m}{E} \right) \right),
\label{akth}
 \end{aligned}
\end{equation}
where $m$ is the mass term defined by Eqs.\eqref{defns} and \eqref{masses}, $E$ represents the eigenvalues of the Hamiltonian given by Eq.\eqref{evals}, and $z$ and $\delta$ are given by Eqs. \eqref{z} and \eqref{delta} respectively. We have dropped the indices $^{\alpha,n}$ to aid the notation throughout the rest of this paper.}

{Since $A_k$ along with the coefficients a and b are $2 \pi$ periodic in $\theta$, we find that the Chern number for the anisotropic DPs reduces to
\begin{equation}
\nu^{\alpha,n} = \frac{1}{2\pi} \int^{2 \pi}_0  \left[A_\theta \right]^\infty_0  d \theta.
\label{nu}
\end{equation}}

{The values of $A_\theta$ for $k=0$ and $k=\infty$ are given by
\begin{equation}
\begin{aligned}
\lim_{k \rightarrow \infty} A_\theta  = & - \frac{1}{2}   \frac{\partial \delta}{\partial \theta}    \\
  + \frac{1}{2}   \frac{\partial \delta}{\partial \theta}   |z| & \left( a^*(\infty,\theta) b(\infty,\theta) + h.c. \right), \\
\lim_{k \rightarrow 0} A_\theta  =  & - \frac{1}{2}   \frac{\partial \delta}{\partial \theta} |a(0,\theta)|^2  \left( 1 - \text{sgn}(m)  \right)  \\
&  - \frac{1}{2}   \frac{\partial \delta}{\partial \theta} |b(0,\theta)|^2   \left( 1 + \text{sgn}(m)   \right) \\
= & 
\begin{cases}
 -  \frac{\partial \delta}{\partial \theta} |b(0,\theta)|^2 & m > 0, \\
-  \frac{\partial \delta}{\partial \theta} |a(0,\theta)|^2 & m < 0. \\
\end{cases}.
\end{aligned}
\end{equation}}

{Thus the pre-quench Chern number for the anisotropic DPs becomes 
\begin{equation}
\nu^{\alpha,n} =
\begin{cases}
\frac{\alpha}{2} \text{sgn}(m^{\alpha,n}) & \text{for the ground state} \\
-\frac{\alpha}{2} \text{sgn}(m^{\alpha,n}) & \text{for the excited state.}
\end{cases}
\label{caniso}
\end{equation}}

{We note that this contribution to the Chern number is the same for each anisotropic SDP;  therefore the total contribution from all the SDPs can simply be written as
\begin{equation}
\nu^{aniso} = \frac{3}{2} \left( \text{sgn }(m^{+}) - \text{sgn }(m^{-}) \right).
\end{equation}}

{For an isotropic DP, the values of $A_k$ and $A_\theta$ are once more given by Eq.\eqref{akth}, but with $m$ replaced by $\mathcal{M}$ (given by Eq.\eqref{masses}) and with $c_x$ and $c_y$ replaced by values given by Eq. \eqref{isodefs}). We note here that the dependence of $c_x$ on the constant $c$ is now with opposite sign, thus for the isotropic DPs we have curvatures
for opposite signs. Since the coefficients $a$, $b$ and thus $A_k$ and $A_\theta$ are all independent of $\theta$ in this case, the Chern number is simply}
\begin{equation}
\nu = \left[ A_\theta \right]^{\infty}_0,
\end{equation}
with
{
\begin{equation}
\begin{aligned}
 \lim_{k\rightarrow \infty} A_\theta   & =  -\frac{\alpha}{2} \\
&+ \frac{\alpha}{2}  \left( a^*(\infty,\theta) b(\infty,\theta) + h.c. \right) , \\
  \lim_{k\rightarrow 0}  A_\theta  &  = -\frac{\alpha}{2} |a(0,\theta)|^2 \left(1-\text{sgn}(m)\right) \\
& -\frac{\alpha}{2}  |b(0,\theta)|^2 \left(1 + \text{sgn}(m)\right) .
\end{aligned}
\end{equation}}

{The contributions to the total pre-quench Chern number for the ground state of the system from both isotropic DPs are thus
\begin{equation}
\nu^{iso}= \frac{1}{2} \left( \text{sgn }(m^-) - \text{sgn }(m^+) \right).
\label{ciso}
\end{equation}}

{We note here the difference in chirality between the isotropic DP \eqref{caniso} and its surrounding anisotropic DPs \eqref{ciso}}.

\section{Chern number after the quench}
\label{postquapp}

{
Performing a sudden quench on the ground state of the system from a state $(M,\phi)$ to a state $(M',\phi')$ we use Eq.\eqref{gs} to obtain the following relationships
\begin{equation}
\begin{aligned}
a & = f_- f'_- + f_+ f'_+ , \\
b & =f_+ f'_- - f_- f'_+,
\label{eq_ab}
\end{aligned}
\end{equation}}
where $f_\pm$ is given by Eq.\eqref{f}.

{Next we evolve this new post-quench state with time as given by {Eq.\eqref{timeev}, and finally compute the Chern number for this new time evolved state. We obtain $A_k$ and $A_\theta$} in this case as}
\begin{equation}
A_{k} =  g  \frac{\partial E'}{\partial k}, \qquad A_\theta=   g  \frac{\partial E'}{\partial \theta} + h \frac{\partial \delta}{\partial \theta}, 
\end{equation}
{where the function $g(k,\theta)$ is given by
\begin{equation}
\begin{aligned}
g(k, \theta) & = t \left(b^2 - a^2 \right) - \frac{m' a b}{E' k |z| } \sin 2  E' t.
\label{gh}
\end{aligned}
\end{equation}
and the function $h(k,\theta)$ by Eq.\eqref{h}.}

\end{appendices}


\begin{thebibliography}{10}

\bibitem{chaikin95}
P.M. Chaikin and T.C. Lubensky
\newblock{\em Principles of Condensed Matter Physics}
\newblock (Cambridge University Press, Cambridge, UK, 1995).

\bibitem{cardy} 
J. Cardy
\newblock {\em Scaling and Renormalization in Statistical Physics} (Cambridge University Press, Cambridge, UK, 1996).

\bibitem{sachdev}
S. Sachdev
\newblock {\em Quantum Phase Transitions }
\newblock (Cambridge University Press, Cambridge, UK, 2007).

\bibitem{duttaetal}
 A. Dutta, G. Aeppli, B. K. Chakrabarti, U. Divakaran, T. F. Rosenbaum, and D. Sen
 \newblock{\em Quantum Phase Transitions in Transverse Field Spin Models: From Statistical Physics to Quantum Information}
 \newblock (Cambridge University Press, Cambridge, UK, 2015).
 
 \bibitem{goerbig09}
M. O. Goerbig,
\newblock Lecture notes for the Singapore session ``Ultracold Gases and
Quantum Information'' of Les Houches Summer School, 2009

\bibitem{h}
M. Haldane
\newblock Phys. Rev. Lett. {\bf 61}, 18, (1988).


\bibitem{hk}
M. Z. Hasan and C. L. Kane,
\newblock Rev. Mod. Phys. 82, 3045, 2010.

\bibitem{moore}
J. E. Moore
\newblock Nature, 464, 194-198, 2010.

\bibitem{Qi10} X. Qi and S. Zhang, Physics Today {\bf 63}, 33 (2010).
\bibitem{Qi11}  Qi and S. Zhang, Rev. Mod. Phys. {\bf 83} 1057 (2011).

\bibitem{bernvig13} B. A. Bernevig with T. L. Hughes, {\it Topological Insulators and Topological Superconductors}, Princeton University Press, Princeton and Oxford (2013).




 \bibitem{jotzu14} Gregor Jotzu ,	Michael Messer,	Rémi Desbuquois,	Martin Lebrat,	Thomas Uehlinger,	Daniel Greif and  Tilman Esslinger,     Nature
    {\bf 515},
    237  (2014).
    
    \bibitem{garnerone09} S. Garnerone, D. Abasto, S. Haas, and P. Zanardi, Phys. Rev. A {\bf 79}, 032302   (2009).
    
\bibitem{mukherjee12} V Mukherjee, A Dutta, D Sen, Phys. Rev.  B {\bf 85},  024301 (2012).

\bibitem{Patel13_marg}  A. A. Patel, S. Sharma, A. Dutta, {\bf 102}, 46001   (2013).

\bibitem{bermudez09} A. Bermudez, D. Patanè,  L. Amico, M. A. Martin-Delgado,  Phys. Rev. Lett. {\bf 102}, 135702, (2009).

\bibitem{Patel13}  A. A. Patel, S. Sharma, A. Dutta, Eur. Phys. Jour. B {\bf 86}, 1 (2013).



\bibitem{rajak14}
A. Rajak, A. Dutta,  Phys. Rev. E 89, 042125, 2014.


\bibitem{sacra1}
P. D. Sacramento,
 Phys. Rev. E {\bf 90} 032138, (2014).


\bibitem{ccb}
M. D. Caio, N. R. Cooper and M. J. Bhaseen,
Phys. Rev. Lett. {\bf 115}, 236403 (2015).

\bibitem{privitera15} L.  Privitera and  G.  E. Santoro,  arXiv:1508.01883 (2015).

\bibitem{sacramento16}
P. D. Sacramento,  arXiv:1601.05476v1 (2016).

\bibitem{kane05} C. L. Kane and E. J. Mele, Phys. Rev. Lett. {\bf 95}, 226801 (2005).

\bibitem{bernevig06} B. Bernevig, T. Hughes, and S. Zhang, Science {\bf 314}, 1757 (2006).

\bibitem{zhangetal}
H. Zhang, C.-X. Liu, X.-L. Qi, X. Dai, Z. Fang and S.-C. Zhang
\newblock Nat. Phys. {\bf 5}, 438-442, (2009).

\bibitem{fu08}
 L. Fu and C. L. Kane, Phys. Rev. Lett. {\bf 100}, 096407 (2008).
 
\bibitem{fkm}
L. Fu, C. L. Kane and E. J. Mele,
Phys. Rev. Lett. {\bf 98}, 106803 (2007).

\bibitem{kitaev09} Alexei Kitaev, Chris Laumann, arXiv:0904.2771 (2009), Les Houches Summer School \textit{Exact methods in low-dimensional physics and quantum computing.} 


\bibitem{calabrese}  P. Calabrese and J. Cardy, Phys. Rev. Lett.  {\bf 96}  136801 (2006).


\bibitem{PolkovnikovRev} A. Polkovnikov, K. Sengupta, A. Silva, and
	M. Vengalattore, Rev. Mod. Phys. {\bf 83}, 863 (2011).	
	
\bibitem {dziarmaga10} J. Dziarmaga, Advances in Physics  {\bf 59}, 1063 (2010).


\bibitem{bdp}
M. Bukov, L. D'Alessio and A. Polkovnikov,
Adv. Phys. {\bf 64}  (2016).

\bibitem{oka09}T. Oka and H. Aoki
Phys. Rev. B {\bf 79}, 081406(R) (2009).

\bibitem{inouetan}
J.-I. Inoue and A. Tanaka,
Phys. Rev. Lett. {\bf 105}, 017401, (2010).

\bibitem{kitagawa10} T. Kitagawa, E. Berg, M. Rudner, and E. Demler, Phys. Rev. B
{\bf 82}, 235114 (2010).

\bibitem{dom1}
H. Dehghani, T. Oka and A. Mitra,
Phys. Rev. B. 90, 195429, (2014).

\bibitem{dom2}
H. Dehghani, T. Oka and A. Mitra,
Phys. Rev. B. {\bf 91}, 155422, (2015).

\bibitem{heyl13} M. Heyl, A. Polkovnikov, and S. Kehrein, Phys. Rev. Lett., {\bf 110}, 135704 (2013).

\bibitem{sharma15} S. Sharma, S. Suzuki and A. Dutta, Phys. Rev. B {\bf 92}, 104306 (2015).

\bibitem{sharma16} S, Sharma, U. Divakaran, A. Polkovnikov, and A. Dutta,
Phys. Rev. B {\bf 93}, 144306 (2016); U. Divakaran, S, Sharma, and A. Dutta,
arXiv:1601.04851 (2016), Phys. Rev. E (in press). 

\bibitem{gambassi11} A. Gambassi and  A.  Silva, arXiv: 1106.2671 (2011); P. Smacchia and A. Silva, Phys. Rev. E {\bf 88}, 042109, (2013).

\bibitem{russomanno15} A. Russomanno, S. Sharma, A. Dutta and G. E. Santoro, J. Stat. Mech., P08030  (2015).


\bibitem{dorner12} R. Dorner, J. Goold, C. Cormick, M. Paternostro and V. Vedral, Phys. Rev. Lett. {\bf 109}, 160601 (2012).

\bibitem{sharma15_PRE} S. Sharma and A. Dutta, Phys. Rev. E {\bf 92}, 022108 (2015).


\bibitem{lrg}
N. H. Lindner, G. Refael and V. Galitski,
 Nat. Phys. {\bf 7}, 490-495, (2011). 
 
 \bibitem{cdsm}
J. Cayssol, B. Dora, F. Simon and R. Moessner,
 Rapid Res. Letts. {\bf 7} 101 (2013).
 
 \bibitem{thakurathi13} M. Thakurathi, A. A. Patel, D. Sen, and A. Dutta, Phys. Rev. B 88, 155133 (2013).
 
 \bibitem{thakurathi14} M. Thakurathi, K. Sengupta and D. Sen, Phys. Rev. B {\bf 89}, 235434 (2014).
 
 \bibitem{mukherjee09} V. Mukherjee and A. Dutta, J. Stat. Mech. (2009) P05005.
 
 \bibitem{das10} A. Das, Phys. Rev. B {\bf 82}, 172402 (2010).
 
\bibitem{quan06} H.T. Quan, Z. Song, X.F. Liu, P. Zanardi, and C.P. Sun, Phys.Rev.Lett. {\bf 96}, 140604 (2006).
 
\bibitem{sharma12}
S. Sharma, V. Mukherjee, A. Dutta,
Eur. Phys. Jour.  B {\bf 85}  1, (2012).

\bibitem{mukherjee12a}
V Mukherjee, S Sharma, A Dutta
Phys. Rev.  B {\bf 86}  020301 (2012).

\bibitem{nag12} T. Nag, U. Divakaran and A. Dutta, 	Phys. Rev. B {\bf 86}, 020401 (R) (2012); S. Suzuki, T. Nag, and A. Dutta, Phys. Rev. A {\bf 93}, 012112 (2016).
 
 \bibitem{russomanno12} A. Russomanno, A. Silva, and G. E. Santoro Phys. Rev. Lett. {\bf 109}, 257201 (2012).
 
 \bibitem{sharma14} S. Sharma, A. Russomanno, G. E. Santoro and A. Dutta, EPL {\bf 106},  67003 (2014).

\bibitem{nag14}T. Nag, S. Roy, A. Dutta,and D. Sen, Phys. Rev.  B {\bf 89}, 165425 (2014); T. Nag, D. Sen, and A. Dutta, Phys. Rev. A {\bf 91}, 063607 (2015). 

\bibitem{agarwala16}
A. Agarwala, U. Bhattacharya, A. Dutta, D. Sen,
 Phys. Rev. B {\bf 93}, 174301, 2016.
 
 \bibitem{lazarides15} A. Lazarides, A. Das and  R. Moessner,  Phys. Rev. Lett. {\bf 115}, 030402 (2015).

\bibitem{hh}
G. B. Hal\'asz and A. Hamma,
 Phys. Rev. Letts. {\bf 110}, 170605 (2013).

\bibitem{alessio15} D' Alessio and M. Rigol, Nat. Comm. {\bf 6}, 8336 (2015).

\bibitem{fdgy1}
M. S. Foster, M. Dzero, V. Gurarie and E. A. Yuzbashyan,
Phys. Rev. B. {\bf 88}, 104511, (2013).

\bibitem{fdgy2}
M. S. Foster, M. Dzero, V. Gurarie and E. A. Yuzbashyan,
Phys. Rev. Lett. {\bf 113}, 076403 (2014).


\bibitem{sp}
D. Sticlet and F. Pi\'echon,
Phys. Rev. B. {\bf 87}, 115402 (2013).

\bibitem{bs}
C. Bena and L. Simon, 
Phys. Rev. B. {\bf 83}, 115404 (2011). 






\end{thebibliography}
\end{document}